\documentclass[12pt,article]{article}
\usepackage[T2A]{fontenc}
\usepackage[utf8]{inputenc}
\usepackage{amsthm,amsfonts,amssymb,amscd,cite}
\usepackage[english]{babel}
\usepackage{graphicx}
\usepackage[tbtags]{amsmath}

\DeclareSymbolFont{bletters}{OML}{cmm}{bx}{it}
\DeclareMathSymbol{\bla}{\mathord}{bletters}{'025}
\DeclareMathSymbol{\bmu}{\mathord}{bletters}{'026}
\DeclareMathSymbol{\bnu}{\mathord}{bletters}{'027}
\DeclareMathSymbol{\bth}{\mathord}{bletters}{'022}
\DeclareMathSymbol{\bfI}{\mathord}{bletters}{"49}
\DeclareMathSymbol{\bdl}{\mathord}{bletters}{"0E}
\DeclareMathSymbol{\bDl}{\mathord}{bletters}{"001}
\def \bpi{\boldsymbol\pi}

\def \bom{\boldsymbol\omega}

\def \si{\sigma}

\def \BZ{\mathbb{Z}}

\def \tr{{\rm tr}\,}

\begin{document}

\title {
$${}$$\\
{\bf The partition function of the four-vertex model in inhomogeneous external \\ field and trace statistics
}}
\author{
$${}$$
{\bf Nikolay Bogoliubov, Cyril Malyshev} \\
$${}$$\\
{\it St.-Petersburg Department of Steklov Institute of Mathematics, RAS}\\
{\it Fontanka 27, St.-Petersburg,
RUSSIA}\\
}

\date{}
\maketitle

\vskip0.5cm
\begin{abstract}
\noindent
The exactly solvable four-vertex model with the fixed boundary conditions in the presence of inhomogeneous linearly growing external field is considered. The partition function of the model is calculated and represented in the determinantal form. The established connection with the boxed plane partitions allows us to calculate
the generating function of plane partitions with the fixed sums of their diagonals.
The obtained results are another example of the connection
of integrable models with the enumerative combinatorics.
\end{abstract}

\thispagestyle{empty}
\newpage

\section{Introduction}

The vertex models with fixed boundary conditions of two-dimensional statistical mechanics play an important role in contemporary studies of integrable models \cite{kor, iz, z, bpz, bofv, cp1, kp}.
There is an intriguing connection of these models with enumerative combinatorics \cite{stan1, kup, bres, bmumn}, the theory of symmetric functions \cite{macd, koh}, and with the limit shapes phenomena \cite{resh, lyb, cp3, cp2, gier}.

The four-vertex model is a particular case of the six-vertex model \cite{bax}
in which two vertices are frozen out. The Quantum Inverse Scattering Method (QISM)  \cite{fad, kbi} was applied to the solution of the four-vertex model on a finite lattice with different boundary conditions  in \cite{bt,bj}. The partition function of the
model was calculated, and in the case of the fixed boundary conditions it was represented in the determinantal form. In the cited papers the connection of
the model with the theory of random lattice paths and plane partitions \cite{stan1, bres} was discussed. The relation of the model with the infinite anisotropy limit of the $XXZ$ Heisenberg magnetic chain was studied in \cite{bmumn}.

Non-intersecting lattice paths and determinants are fundamental tools for analyzing
plane partitions. In the paper \cite{st} Stanley derived the
norm-trace generating function -- the generating function for plane partition with unbounded parts and with the fixed sum
of its main diagonal. Later the trace formula was generalized in \cite{gan} on the $l$-traces case. The trace statistics
of the bounded plane partitions (boxed plane partitions) equipped with a special weights was considered in \cite{kam, tri},
where the generating functions of these partitions were expressed as triple products.

Computing partition functions of two-dimensional lattice models is one of the central problems in statistical mechanics.
In the present paper we consider the four-vertex model on the finite lattice in presence of an inhomogeneous linearly growing external field. The fixed boundary conditions are applied. Using QISM we have calculated the partition function of this model and represent it as determinant. In the case of the unlimited height of the box the determinant may be calculated
and the partition function is expressed as double product.
Exploring the connection of the considered vertex model with the boxed plane partitions
we have represented in the determinantal form the generation functions of plane partitions with the fixed sums of their diagonals.
In the case of unbounded plane partitions the resulting formulas
generalize the trace generating functions by Stanley and  by Gansner \cite{st, gan}.

The paper is organized as follows. In section 2 we define the anisotropic four-vertex model with the fixed boundary conditions on a two dimensional square lattice of finite size. The inhomogeneous external field is introduced
and the partition function of the model is considered. In section 3 the spin description of the model
that allows us to apply the QISM to the solution of the model is discussed. The partition function of the model
is calculated and expressed as the determinant in section 4. The obtained determinant is calculated in the case of semi-infinite lattice and represented in a double product form. The connection of the model with the boxed plane partitions with the fixed sums of their diagonals is established in section 5. This connection allows us
to study the trace statistics of plane partitions. Section 6 is the conclusion.

\section{Four-vertex model}

Consider a square grid of $2N$ vertical lines and $M+1$ horizontal ones. A four-vertex model is described by four different arrows
arrangements pointing in and out of each vertex on a grid (Fig.~1). Representing the arrows pointed up and to the right by the lines we obtain an alternate description of the vertices in terms of lines flowing through the vertices. Since a lattice edge can exist
in two states, line or no line, there exists a one-to-one
correspondence between the arrow configuration on the lattice and
the graphs of lines on the lattice -- \textit{nests of lattice paths}.
\begin{figure}[h]
\centering
\includegraphics{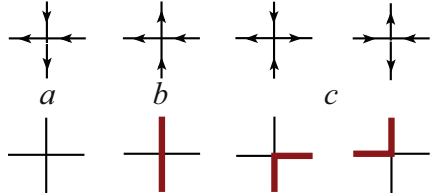}
\caption{The four allowed types of vertices in terms of arrows and lines.}
\end{figure}
A statistical weight corresponds to each type of the vertices and there are four
vertex Boltzmann weights $\omega_a$ , $\omega_b$ and $\omega_c$.
For the general inhomogeneous case the weights are site dependent.

The allowed configurations of arrows depend on the
imposed boundary conditions which are specified by the direction
of the arrows on the boundary of the grid. Under the fixed boundary conditions we shall understand the following arrangement of boundary arrows: the arrows on the top and bottom of the $N$ vertical lines (counting from the left) are pointing inwards, and the arrows on the top and bottom of the last $N$ ones are pointing outwards. All arrows on the left and right boundaries of the grid
are pointing to the left.

To enumerate all possible configurations $\{\nu\}$ of the vertices  it is more convenient to use the
description of the model in terms of flowing lines and to represent the allowed configurations
as nests of lattice paths. The path is running from one of the $N$ down left vertices to
the top $N$ right ones and always moves east or north. The paths cannot touch each other
and arbitrary number of consequent steps are allowed in vertical direction while only a single
step is allowed in the horizontal one. The length of the path is $N+M$ and there are $N$ paths in a nest of lattice paths. A typical nest of lattice paths is represented in Fig.~2.
\begin{figure}[h]
\centering
\includegraphics{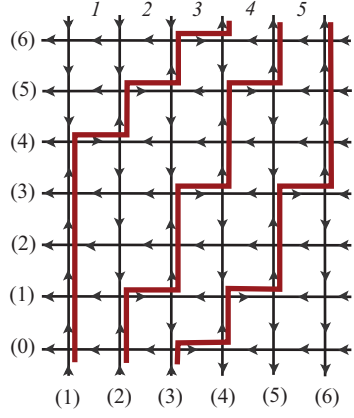}
\caption{A typical nest of admissible lattice paths with
the fixed boundary conditions.}
\end{figure}
The step in the horizontal direction is defined by the neighbouring pairs of vertices ($c$). The number of these pairs in the admissible path is $N$ and, hence, the number of ($c$) vertices in the nest of lattice paths is $l^c = 2N^2$.
Since only one step is allowed in horizontal direction the number of ($b$) vertices in the path
is $M-N+1$. It means that the numbers of ($a$) and ($b$)  vertices in the nest of lattice paths are equal: $l^a=l^b=N(M-N+1)$.

In our paper we shall consider the inhomogeneous model defined by site depending weights $(\bom_a)_{jk}$, $(\bom_b)_{jk}$, $(\bom_c)_{jk}$
($j=1, 2, \ldots, 2N$, $k=0, 1, \ldots, M$). To define these weights it is convenient to label by $s$ the columns of the grid formed by the horizontal edges lying between the neighbouring vertical lines $s$ and $s+1$
($s=1, 2, \ldots, 2N-1$).
The coordinates $\mu_k^s$ of arrows turned to the right in the $s^{\rm th}$ column form a strict partition ${\bmu}^s$, that is $M\geq \mu_1^s > \mu_2^s > \ldots > \mu_K^s \geq 0$ $(K\leq N)$,
with the parts satisfying the condition $\mu_k^s > \mu_{k+1}^s+1$. The norm of the partition is equal to the sum of its parts $|\bmu^s|=\sum_k \mu^s_k$.
The nest of lattice paths $\nu$ uniquely defines the configuration of
vertex weights on a lattice and also the set of coordinates of arrows turned to the right $\bmu \equiv (\bmu^1, \bmu^2, \ldots , \bmu^{2N-1})$.

The vertices ($c$) appear in configurations $\{\nu \}$ only in pairs,
since a single step is allowed in horizontal direction (Fig.~2). We define their Boltzmann weights as
\begin{equation}
\label{cwe}
(\bom_c)_{s,\mu_k^s} =(\bom_c)_{s+1,\mu_k^s}= \exp{( h_s \mu_k^s )}\,,\quad s=1, 2, \ldots, 2N-1\,
\end{equation}
so that
\begin{equation}
\label{cwe1}
(\bom_c)_{s,\mu_k^s}\times (\bom_c)_{s+1,\mu_k^s}= \exp{(2 \mu_k^s h_s)}\,,
\end{equation}
where $2 h_s\equiv -\beta \mathcal{H}_s$, $\beta$ is the inverse temperature, and $\mathcal{H}_s$ is the external horizontal field.
Equation (\ref{cwe1}) may be considered as the weight of an arrow turned to the right.
The weights of vertices ($a$) and ($b$) are inhomogeneous only in
horizontal direction:
\begin{equation}\label{wei}
(\bom_a)_{jk} = e^{-\beta\left( \mathcal{E}_a-\mathcal{V}_j \right)}=e^{-\beta \mathcal{E}_a}\omega^{-1}_j\,,
\qquad (\bom_b)_{jk} = e^{-\beta\left(
\mathcal{E}_b + \mathcal{V}_j\right)} = e^{-\beta \mathcal{E}_b}\omega_j\,,
\end{equation}
where $\mathcal{E}_a$, $\mathcal{E}_b$ are the energies of the vertices ($a$) and ($b$), and $\mathcal{V}_j\geq 0$ is the external vertical field, $j = 1, 2, \ldots, 2N$.

The partition function of the  model introduced above is equal to
\begin{multline}
Z(\bom_a, \bom_b, \bom_c) = e^{-\beta N(M-N+1)(\mathcal{E}_a +\mathcal{E}_b)}
\sum_{\{\nu\}} \prod_{j=1}^{2N} \omega_j^{l_j^b-l_j^a} \prod_{k=1}^K(\bom_c)_{j,\mu_k^s}\\
\equiv e^{-\beta N(M-N+1)(\mathcal{E}_a +\mathcal{E}_b)}\,
\widetilde{Z} (\bom, \bom_c)\,,  \label{ihpartff}
\end{multline}
where $2N$-tuple $\bom\equiv (\omega_1, \omega_2, \ldots, \omega_{2 N})$ is introduced in right-hand side.
The summation in (\ref{ihpartff}) is taken over all allowed vertex configurations $\nu$ (admissible nests of lattice paths) and
$l_j^a$, $l_j^b$ are the numbers of corresponding vertices in the vertical lines $j$. The modified partition function
$\widetilde{Z} (\bom, \bom_c)$ (\ref{ihpartff}) may be rewritten under the parametrization (\ref{cwe1}) in the form
\begin{equation}\label{gfu}
\widetilde{Z} (\bom, e^{2 \bf h})=\sum_{\{\nu\}} \prod_{s=1}^{2N-1} e^{2|\bmu^s|h_s}  \prod_{j=1}^{2N} \omega_j^{l_j^b-l_j^a} \,,
\end{equation}
where $(2N-1)$-tuple $e^{2 \bf h}\equiv (e^{2 h_1}, e^{2 h_2}, \ldots, e^{2 h_{2N-1}})$ is
introduced to point out that $\widetilde{Z}$ depends now on the parameters $h_s$, $s=1, 2, \ldots, 2N-1$. Two particular cases of the partition
function (\ref{gfu}) are of importance below. When ``horizonal'' field is absent, $h_s=0$, $s=1, 2, \ldots, 2N-1$, we put:
\begin{equation}\label{gfu1}
\widetilde{Z}_{\cal V} (\bom) \equiv \sum_{\{\nu\}} \prod_{j=1}^{2N} \omega_j^{l_j^b-l_j^a} \,.
\end{equation}
When ``vertical'' field is absent, $\omega_j=1$, $j=1, 2, \ldots, 2N$, we put:
\begin{equation}\label{gfu2}
\widetilde{Z}_{\cal H} (e^{2 \bf h}) \equiv \sum_{\{\nu\}}
\prod_{s=1}^{2N-1} e^{2|\bmu^s|h_s}\,.
\end{equation}
In the absence of the external fields $\mathcal{E}$ and $\mathcal{H}$ the modified partition function (\ref{gfu}) is equal to the number of lattice paths on a $2N\times(M+1)$ square grid described  above.

\section{Spin description of the model}

To apply the Quantum Inverse Scattering Method to the solution of the model we shall use the spin description of the model.
With each vertical and horizontal bond of the grid we associate
a local space isomorphic to $\mathbb{C}^2$, with spin ``up''
$\left(
\begin{array}{c}
1 \\
0
\end{array}
\right)$
and spin ``down''
$\left(
\begin{array}{c}
0 \\
1
\end{array}
\right)$
states forming a natural basis in this space. The spin ``up'' state on the vertical bond corresponds to the arrow pointing up, while the spin ``down'' state to the arrow pointing down. The spin ``up'' state on the
$i^{\rm th}$ horizontal bond $\left(
\begin{array}{c}
1 \\
0
\end{array}
\right)_i\equiv |\leftarrow \rangle_i$ corresponds to the
horizontal arrow
pointing to the left, spin ``down'' state to the arrow pointing to the right $\left(
\begin{array}{c}
0 \\
1
\end{array}
\right)_i\equiv |\rightarrow \rangle_i$. States corresponding to the $l^{\rm th}$ vertical line are denoted by arrows directed upwards or downwards:
$\left(
\begin{array}{c}
1 \\
0
\end{array}
\right)_l \equiv |\uparrow \rangle_l$ and
$\left(
\begin{array}{c}
0 \\
1
\end{array}
\right)_l\equiv |\downarrow \rangle_l$.
The auxiliary space $\mathbb{V}$ is the tensor product of all the local spaces associated with the vertical lines:
$\mathbb{V}=(\mathbb{C}^2)^{\otimes 2N}$, and the quantum space $\mathbb{H}$ is the tensor product of all local spaces associated with the horizontal lines: $\mathbb{H} = (\mathbb{C}^2)^{\otimes (M+1)}$.

The partition function (\ref{gfu}) will be calculated by the Quantum Inverse Scattering Method \cite{fad}. With each
vertex of the lattice we associate an operator acting in the space $\mathbb{C}^2\otimes \mathbb{H}$. This operator is called $L$-operator and it acts trivially (as the identity operator) on all sites except the fixed vertex.
The $L$-operator of the four vertex model is equal to \cite{bt, bj}:
\begin{equation}\label{lop}
L(n|u)=-u \check{e} \check{e}_n+u^{-1} \hat{e}\check{e}_n+ \sigma^{+}\sigma_n^{-}+
\sigma^{-}\sigma_n^{+}
=\left(
\begin{array}{cc}
-u\check{e}_n & \sigma_n^{-} \\
\sigma_n^{+} & u^{-1}\check{e}_n
\end{array}
\right)\,,
\end{equation}
where the parameter $u\in \mathbb{C}$, $\sigma^{z,\pm }$ are the Pauli matrices, and $\check{e}=\frac 12(1+\sigma^z)$, $\hat{e}=\frac 12(1-\sigma^z)$ are  projectors on the states with spins ``up'' and
``down'' respectively. 
The matrix with subindex $n$ acts
nontrivially only in the $n^{\rm th}$ space: ${\sf s}_n=1\otimes
\cdots\otimes
1\otimes \underbrace{{\sf s}}_n\otimes 1\otimes \cdots \otimes 1$, where
$0\leq n\leq M$.
The monodromy matrix is the product of $L$-operators
\begin{equation}
T(u)=L(M|u) L(M-1|u) \cdots L(0|u)=\left(
\begin{array}{cc}
A(u) & B(u) \\
C(u) & D(u)
\end{array}
\right)\,.
\label{mm}
\end{equation}
The matrix elements of the introduced $L$-operator (\ref{lop}) can be represented graphically as dots with the attached arrows (see Fig.~3).
\begin{figure}[h]
\centering
\includegraphics{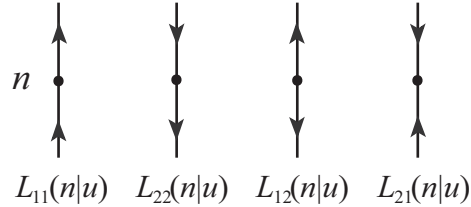}
\caption{Vertex representation of the matrix elements of the
$L$-operator.}
\end{figure}
In a representation of the matrix element $L_{11}(n|u)$ a dot stands for the operator $-u\check{e}_n$, this operator acts on the local spin state, and the only
non-zero matrix element of this operator is $_n\langle \leftarrow
|-u\check{e}_n| \leftarrow \rangle_n$ what gives the vertex $(b)$ (Fig.~1) with a weight $-u$.
In the representation of the  matrix element $L_{22}(n|u)$
a dot is the operator
$u^{-1}\check{e}_n$, and the non-zero matrix element $_n\langle \leftarrow
|u^{-1}\check{e}_n|\leftarrow \rangle_n$ is the
vertex $(a)$ (Fig.~1) with a weight $u^{-1}$.
The nonzero matrix elements $_n\langle \rightarrow |\sigma_n^{-}|\leftarrow \rangle_n$
and $_n\langle \leftarrow |\sigma_n^{+}|\rightarrow \rangle_n$ of
$L_{12} (n|u)$ and $L_{21} (n|u)$ are vertices  $(c)$
(Fig.~1).

The entries of the monodromy matrix (\ref{mm}) are
expressed as  sums over all possible configurations of arrows with different boundary conditions on a one-dimensional lattice with $M+1$ sites (Fig.~4). Namely, the operator $B(u)$ corresponds to the boundary conditions, when
arrows on the top and bottom of the lattice are pointing outwards. Operator $C(u)$ corresponds to the boundary conditions, when arrows on the top and bottom of the lattice are pointing inward. Operators $A(u)$ and $D(u)$
correspond to the boundary conditions, when arrows on the top and bottom of the lattice are pointing up and down respectively.
\begin{figure}[h]
\centering
\includegraphics{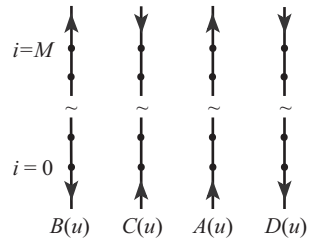}
\caption{Graphic representation of the entries of the monodromy
matrix}
\end{figure}

The operator-valued matrices (\ref{lop}) and (\ref{mm}) are associated with the following $R$-matrix \cite{bt}:
\begin{equation}
R(u, v)=\left(
\begin{array}{cccc}
f(v,u) & 0 & 0 & 0 \\
0 & g(v,u) & 1 & 0 \\
0 & 0 & g(v,u) & 0 \\
0 & 0 & 0 & f(v,u)
\end{array}
\right)\,,  \label{r}
\end{equation}
where
\begin{equation}
f(v,u)=\frac{u^2}{u^2-v^2}\,,\qquad g(v,u)=\frac{uv}{u^2-v^2}.
\label{fg}
\end{equation}
The commutation relations of the matrix elements of the monodromy
matrix (\ref{mm}) are defined by the introduced $R$-matrix (\ref{r}).
The most important relations are:
\begin{equation}
\begin{array}{l}
C(u)B(v)\,=\,g(u,v)\big\{ A(u)D(v)-A(v)D(u)\big\}\,,
\\ [0.2cm]
A(u)B(v)\,=\,f(u,v)\, B(v)A(u)+g(v,u)\, B(u)A(v)\,,  \\
[0.2cm]
D(u)B(v)\,=\,f(v,u)\, B(v)D(u)+g(u,v)\, B(u)D(v)\,,   \\
[0.2cm]
\big[ B(u),B(v) \big]\,=\, \big[ C(u), C(v) \big]=0\,.
\end{array}
\label{cbad}
\end{equation}

The $L$-operator (\ref{lop}) satisfies the relations
\begin{equation}\label{sisi}
e^{-2h\sigma_n^z} L(n|u) e^{2h\sigma_n^z} = e^{2h\sigma^z} L(n|u) e^{-2h \sigma^z}\,,
\end{equation}
and
\begin{equation}\label{hl}
e^{-2h\sigma^z} L(n|u)= e^{-h\si^z} L(n|e^{-2h}u) e^{h\si^z}\,.
\end{equation}
From these equations and from the definition of the monodromy matrix (\ref{mm}) it follows that
\begin{equation}
e^{-2h\sum_{j=0}^M \sigma_j^z} T(u)e^{2h\sum_{j=0}^M \sigma_j^z}
=e^{2h\sigma^z} T(u) e^{-2h\sigma^z}\,, \label{sts}
\end{equation}
and
\begin{equation}\label{ht}
e^{-2h\sum_{j=0}^M j \sigma_j^z} T(u) =
e^{2h(M+1) \sigma^z} e^{-h\sigma^z} T(e^{-2h}u) e^{h\si^z} e^{-2h\sum_{j=0}^M j\sigma_j^z}\,.
\end{equation}
The relations
\begin{equation}
\begin{array}{l}
e^{2h\sum_{j=0}^M j\hat{e}_j} B(u) = e^{hM}B(e^{-h}u) e^{2h\sum_{j=0}^M j\hat{e}_j}\,,\\ [0.2cm]
C(u)e^{2h\sum_{j=0}^M j\hat{e}_j} =  e^{hM} e^{2h\sum_{j=0}^M j\hat{e}_j} C(e^{h}u) \,,
\end{array}
\label{b}
\end{equation}
and
\begin{align*}
e^{2h\sum_{j=0}^M j\hat{e}_j} A(u) &= e^{h(M+1)} A(e^{-h}u) e^{2h\sum_{j=0}^M j \hat{e}_j} \,,\\
e^{2h\sum_{j=0}^M j \hat{e}_j} D(u) &= e^{-h(M+1)} D(e^{-h}u) e^{2h\sum_{j=0}^M j \hat{e}_j}
\end{align*}
are the consequence of Eq.~(\ref{ht}).
\begin{figure}[h]
\centering
\includegraphics{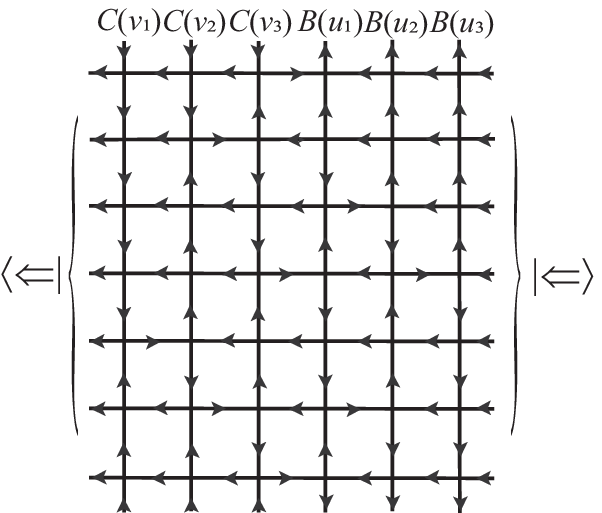}
\caption{One of configurations that contribute into the scalar product}
{$\langle \Leftarrow |C(v_1) C(v_2) C(v_3) B(u_1) B(u_2) B(u_3)|\Leftarrow \rangle$.}
\end{figure}

Let us consider the scalar product
\begin{equation}\label{scpr}
W(\mathbf{v}, \mathbf{u})=\langle \Leftarrow
|C(v_1) C(v_2) \cdots C(v_N) B(u_1) B(u_2) \cdots B(u_N)|\Leftarrow \rangle\,,
\end{equation}
where $\mathbf{v}\equiv (v_1, v_2, \ldots, v_N)$ and $\mathbf{u}\equiv (u_1, u_2, \ldots, u_N)$ are the sets of $N$ independent parameters,
and the state
\begin{equation}
|\Leftarrow \rangle = \otimes_{i=0}^M | \leftarrow \rangle_i = \otimes_{i=0}^M \left(
\begin{array}{c}
1 \\
0
\end{array}
\right)_i\,. \label{gs}
\end{equation}
The graphical representation of the scalar product (\ref{scpr}) in Fig.~5 corresponds to the model with the fixed boundary conditions.

For arbitrary $N$ and $M$ this scalar product is evaluated by
means of the commutation relations (\ref{cbad}) and for the integrable models associated with the $R$-matrix (\ref{r})
may be represented in the
determinantal form \cite{bt, bj}:
\begin{multline}\label{dscpr}
W(\mathbf{v},\mathbf{u})=
\Biggl\{ \prod_{1\leq k < j \leq N}g(v_j,v_k)\prod_{1\leq m <l \leq N} g(u_m,u_l) \Biggr\}
\det \widetilde{H}({\bf v},{\bf u})\\
=\prod_{k=1}^N \left(\frac{v_k}{u_k}\right)^{N-1} \frac{\det \widetilde{H}({\bf v}, {\bf u})}{\Delta_N({\bf v}^2) \Delta_N({\bf u}^{-2})}\,,
\end{multline}
The matrix $\widetilde{H}({\bf v},{\bf u})\equiv \bigl(\widetilde{H}_{km}({\bf v},{\bf u})\bigr)_{1\leq k,m \leq N} $ is given by the entries
\begin{equation}\label{q}
\widetilde{H}_{km}({\bf v},{\bf u}) = \frac{ \alpha_{M+1} (v_m)\delta_{M+1} (u_k) \Big(\displaystyle{\frac{u_k}{v_m}}
\Big)^{N-1} - \alpha_{M+1} (u_k) \delta_{M+1}(v_m)
\Big(\frac{u_k}{v_m}\Big)^{-N+1}}
{\displaystyle{\frac{u_k}{v_m}}
-\Big( \frac{u_k}{v_m} \Big)^{-1}}\,,
\end{equation}
where $\alpha_{M+1}(u)$ and $\delta_{M+1}(u)$ are the eigenvalues of operators $A(u)$ and $D(u)$, Eq.(\ref{mm}):
\begin{align}\label{adv}
A(u)| \Leftarrow \rangle =\alpha_{M+1} (u)|\Leftarrow \rangle \,,\\
D(u)|\Leftarrow \rangle =\delta_{M+1} (u)|\Leftarrow \rangle \nonumber \,.
\end{align}
In (\ref{dscpr}), $\Delta_N({\bf x}^2)$ is the Vandermonde determinant
\begin{equation}
\label{van}
\Delta_N({\bf x}^2)=\prod_{1\leq m<k\leq N}(x_k^2-x_m^2)\,.
\end{equation}
For the considered model $\alpha_{M+1} (u)=(-u)^{M+1}$ and $\delta_{M+1} (u)=u^{-(M+1)}$, and the answer for the scalar product (\ref{scpr}) is the following:
\begin{equation}
\label{scpran}
W(\mathbf{v}, \mathbf{u})\,=\,
\frac{(-1)^{MN}\,\prod\limits_{k=1}^N
\Big(\displaystyle{\frac{v_k}{u_k}}
\Big)^{- (M-2N+2)}}{\Delta_N({\bf v}^{2})\Delta_N({\bf u}^{-2})}\,
\det\!\left[ \frac{1- \Big(
\displaystyle{\frac{v_m}{u_k}}
\Big)^{2 (M-N+2)}}{1- \Big(\displaystyle{\frac{v_m}{u_k}}
\Big)^2}\right]_{1\leq k, m \leq N}\,.
\end{equation}

Let us consider the weighted scalar product
\begin{multline}
\label{cb}
W(\mathbf{v}, \mathbf{u}| \mathbf{h})= \langle \Leftarrow |C(v_1){\hat{\varkappa}}_1 C(v_2){\hat{\varkappa}}_2 \cdots C(v_{N-1}){\hat{\varkappa}}_{N-1} C(v_N){\hat{\varkappa}}_{N}\\
\times\, B(u_1) {\hat{\varkappa}}_{N+1} B(u_2)
{\hat{\varkappa}}_{N+2} \cdots  B(u_{N-1}){\hat{\varkappa}}_{2N-1} B(u_N)|\Leftarrow \rangle \,,
\end{multline}
where operator ${\hat{\varkappa}}_n = e^{2h_n \sum_{j=1}^M j \hat{e}_j}$ is introduced, and ${\bf{h}}=(h_1, h_2, \ldots, h_{2N-1})$.
Moving the operators ${\hat{\varkappa}}_j$ with $j=1, 2, \ldots, N$ to the left and with
$j=N+1, N+2, \ldots, 2N-1$ to the right with the help of Eqs.~(\ref{b}), we obtain:
\begin{multline}\label{pfex}
\kappa^{-M} W(\mathbf{v}, \mathbf{u} | \mathbf{h}) = \langle \Leftarrow |C(e^{\sum_{j=1}^N h_j} v_1)\cdots C(e^{\sum_{j=k}^N h_j} v_k)\cdots C(e^{h_N} v_N) \\
\times\, B(u_1) B(e^{-h_{N+1}} u_2) \cdots B(e^{-\sum_{j=1}^{k-1} h_{N+j}} u_k) \cdots B(e^{-\sum_{j=1}^{N-1} h_{N+j}}u_{N})|\Leftarrow \rangle \\
= \langle \Leftarrow | C(\tilde{v}_1) \cdots C(\tilde{v}_N)
B(\tilde{u}_1) \cdots B(\tilde{u}_N)|\Leftarrow \rangle \,,
\end{multline}
where the property  ${\hat{\varkappa}}_n |\Leftarrow \rangle=|\Leftarrow \rangle$ and  $\langle \Leftarrow |{\hat{\varkappa}}_n=\langle \Leftarrow |$ was taken into account, and the
variables $\tilde{v}_k = e^{\sum_{j=k}^N h_j} v_k$ and $\tilde{u}_k=e^{-\sum_{j=1}^{k-1} h_{N+j}} u_k$ were introduced.
The coefficient $\kappa$ in (\ref{pfex}) is equal to
\begin{equation}\label{kappa}
\kappa = e^{\sum_{j=1}^N jh_j}\, e^{\sum_{j=1}^{N-1}(N-j) h_{N+j}}\,.
\end{equation}

From the equation (\ref{pfex}) it follows that the weighted scalar product (\ref{cb}) is expressed through the scalar product (\ref{scpr})
by the equality:
\begin{equation}\label{rel}
W(\mathbf{v}, \mathbf{u}| \mathbf{h})=\kappa^M W(\tilde{\mathbf{v}}, \tilde{\mathbf{u}})\,.
\end{equation}

\section{Partition function}

The graphical representation of the operators $B(u)$ and $C(u)$ and of the state (\ref{gs}), which describes arrows pointing to the left on the left and right boundaries of the grid,
allows us to interpret the scalar product (\ref{scpr}) as the sum over all allowed
configurations of vertices on a square lattice with the fixed boundary conditions (see Fig.~5):
\begin{multline}
\label{screp}
W(\mathbf{v},\mathbf{u}) = \sum_{\{\nu\}} \prod_{k=1}^{N} (-v_{k})^{l_k^b} (v_{k}^{-1})^{l_k^a}
\prod_{j=1}^N(-u_j)^{l_j^b} (u_j^{-1})^{l_j^a} \\
=(-1)^{MN}\sum_{\{\nu\}} \prod_{k=1}^{N} (v_{k})^{l_k^b} (v_{k}^{-1})^{l_k^a}
\prod_{j=1}^N (u_j)^{l_j^b} (u_j^{-1})^{l_j^a}\,.
\end{multline}
Let us put
\begin{equation}
\label{vwinm}
v_j=\omega_j, \qquad u_j=\omega_{N+j}\,,\qquad
1 \leq j\leq N\,,
\end{equation}
and introduce the notations
\begin{equation}\label{vwinm1}
\bom^{\rm I} \equiv (\omega_1, \omega_2, \ldots, \omega_N)\,,\quad \bom^{\rm II} \equiv
(\omega_{N+1}, \omega_{N+2}, \ldots, \omega_{2N})
\end{equation}
so that $2N$-tuple $\bom=(\bom^{\rm I}, \bom^{\rm II})$. Then, using (\ref{vwinm}) in (\ref{screp}), we find out that the ``vertical'' partition function (\ref{gfu1}) is expressed through the scalar product (\ref{scpr}):
\begin{equation}
\widetilde{Z}_{\cal V} (\bom)\,=\, (-1)^{MN} W(\bom^{\rm I}, \bom^{\rm II})\,,
\label{pfscp}
\end{equation}
where (\ref{vwinm1}) is accounted for.

By the construction, the partition function (\ref{gfu}) in the presence of the external horizontal field ${\bf h}$
and the ``weighted'' scalar product (\ref{cb}) are related:
\begin{equation}\label{gfcb}
\widetilde{Z}(\bom, e^{2 \bf h})=(-1)^{MN} W(\bom^{\rm I}, \bom^{\rm II} | \mathbf{h})\,.
\end{equation}
This relation together with Eqs.~(\ref{scpran}) and (\ref{rel}) solves the problem of calculation of the partition function (\ref{gfu}) and, respectively, of (\ref{ihpartff}).

Let us pass from $N$-tuples ${\bom}^{\rm I}$ and ${\bom}^{\rm II}$ (\ref{vwinm1}) to $N$-tuples $\tilde{\bom}^{\rm I}$ and $\tilde{\bom}^{\rm II}$:
\begin{equation}
\begin{array}{l}
\tilde{\bom}^{\rm I} \equiv (\omega_1\,e^{\sum_{j=1}^N h_j},
\ldots, \omega_k\,e^{\sum_{j=k}^N h_j}, \ldots, \omega_N\, e^{h_N})\,,\\ [0.3cm] \tilde{\bom}^{\rm II} \equiv
(\omega_{N+1},\ldots, \omega_{N+k}\,
e^{-\sum_{j=1}^{k-1} h_{N+j}}, \ldots, \omega_{2N}\, e^{-\sum_{j=1}^{N-1} h_{N+j}})\,.
\end{array}\label{vwinm3}
\end{equation}
Substituting representation (\ref{vwinm3}) into (\ref{scpran}) and (\ref{rel}), we obtain the partition function (\ref{gfu}) in the determinantal form:
\begin{multline}\label{gfua}
\widetilde{Z}(\bom, e^{2 \bf h})=\frac{1}{\Delta_N\Big(\big( \tilde{\bom}^{\rm I}\big)^2\Big) \Delta_N\Big(\big(\tilde{\bom}^{\rm II}\big)^{-2}\Big)}\\
\times\,\prod_{k=1}^N\Big( \frac{\omega_{k}}{\omega_{N+k}} \Big)^{-(M-2N+2)}
\big(e^{\sum_{j=1}^{k-1} h_{N+j}+\sum_{j=k}^N h_j }\big)^{2 (N-1)}\\
\times\det \left[\frac{1- \big(\frac{\omega_m}{\omega_{N+k}} e^{\sum_{j=1}^{k-1} h_{N+j}+\sum_{j=m}^N h_j }\big)^{2(M-N+2)}}
{1-\big( \frac{\omega_m}{\omega_{N+k}} e^{\sum_{j=1}^{k-1} h_{N+j}+\sum_{j=m}^N h_j }\big)^2}\right]_{1\leq k,m \leq N}\,.
\end{multline}
In the derivation of this expression we have used the equality:
\begin{align*}
\prod_{j=1}^N a_j^{jM} \prod_{j=1}^{N-1} a_{N+j}^{(N-j)M}
\prod_{k=1}^N \left(\prod_{j=1}^{k-1} a_{N+j} \prod_{j=k}^{N} a_{j} \right)^{-(M-2N+2)} \\
= \left(\prod_{j=1}^Na_j^{j} \prod_{j=1}^{N-1} a_{N+j}^{(N-j)}\right)^{2(N-1)}\,.
\end{align*}

Let us consider a case when the height of a lattice is large, $M\gg 1$, and its length $N$ satisfies the condition $N\ll M$.
Then the determinant in (\ref{gfua}) turns into the Cauchy determinant and may be calculated. In the considered limit the normalized partition function is equal to
\begin{multline}\label{cda}
\mathcal{Z}(\bom, e^{2\bf h})\,
\equiv\,\lim_{M\rightarrow \infty}\,\, \Biggl\{\prod_{j=1}^N \Bigl(\frac{\omega_{j}}{\omega_{N+j}} \Bigr)^{M}
\widetilde{Z}(\bom , e^{2 \bf h})
\Biggr\} \\
=\,\bigl(e^{\sum_{j=1}^{k-1} h_{N+j}+\sum_{j=k}^N h_j }\bigr)^{2(N-1)}
\prod_{k=1}^N
\prod_{m=1}^N \frac{1}{1-\bigl( \frac{\omega_m}{\omega_{N+k}} e^{\sum_{j=1}^{k-1} h_{N+j}+\sum_{j=m}^N h_j }\bigr)^2}\,.
\end{multline}
The equations (\ref{gfua}) and (\ref{cda}) solve the problem of calculation of the partition function of the considered four-vertex model in general.

Let us study the model in the homogeneous horizontal field
characterized by appropriate $(2N-1)$-tuple $e^{{2 \bf h}_{\rm hom}}\equiv (e^{2h}, e^{2h}, \ldots, e^{2h})$ and consider the case when the vertical field $\mathcal{V}$ is absent. The ``horizontal'' partition function (\ref{gfu2}) takes the form:
\begin{equation}\label{gfuhom}
\widetilde{Z}_{\cal H} (e^{2{\bf h}_{\rm hom}}) = \sum_{\{\nu\}} e^{2h\sum_{s=1}^{2N-1} |\bmu^s|} \,,
\end{equation}
where the sum $\sum_{\{\nu\}}$ is taken over all admissible nests of lattice paths on the $2N\times (M+1)$ grid.

In the limit discussed Eq.~(\ref{gfua}) leads to the determinantal representation alternate to (\ref{gfuhom}):
\begin{multline}\label{gfual}
\widetilde{Z}_{\cal H} (e^{2{\bf h}_{\rm hom}}) = e^{2hN^2(N-1)} \prod_{1\leq m<k\leq N}e^{-2hN} \left(e^{h(k-m)}-e^{-h(k-m)} \right)^{-2}\\
\times\,\det \left[\frac{1- e^{2h(k+m-1)(M-N+2)}}
{1-e^{2h(k+m-1)}} \right]_{1\leq k, m \leq N}\,.
\end{multline}
The determinant in (\ref{gfual}) is calculated by means of the following relation obtained in \cite{kup}:
\begin{equation}\label{kup}
\det \left( \frac{1-s^{j+k-1}}{1-q^{j+k-1}} \right) \\ = \, q^{\frac{N^2(N-1)}{2}} \prod_{1\leq m<k\leq N}\left(q^\frac{k-m}{2}- q^{-\frac{k-m}{2}} \right)^2
\prod_{k=1}^N \prod_{j=1}^N \frac{1-sq^{j-k}}{1-q^{j+k-1}}\,.
\end{equation}
As a result, Eq.~(\ref{gfual}) takes the form
\begin{multline}\label{ydpf}
\widetilde{Z}_{\cal H} (e^{2{\bf h}_{\rm hom}})\,=\,e^{2hN^2(N-1)} \prod_{k=1}^N \prod_{m=1}^N \frac{1-e^{2h(M-N+2+m-k)}} {1-e^{2h(k+m-1)}}\\
=\,e^{2hN^2(N-1)} \prod_{k=1}^N \prod_{m=1}^N \frac{1-e^{2h(M+3-m-k)}}{1 -e^{2h(k+m-1)}}\,.
\end{multline}
The number of admissible nests of lattice paths on the $2N\times (M+1)$ grid appears as $h\rightarrow 0$ in (\ref{ydpf}):
\begin{equation}\label{nspp}
\lim_{h\to 0}\, \widetilde{Z}_{\cal H} (e^{2{\bf h}_{\rm hom}}) = \prod_{k=1}^N \prod_{m=1}^N \frac{M+3-k-m}{k+m-1}\,.
\end{equation}

\section{Plane partitions and generating functions}

A plane partition $\bpi$ is an array
\[
\bpi=\left(
\begin{array}{ccccc}
\pi_{11} & \pi_{12} & \cdots & \pi_{1 j} & \cdots \\
\pi_{21} & \pi_{22} & \cdots & \pi_{2 j} & \cdots\\
\vdots & \vdots &  \ddots & \vdots & \vdots \\
\pi_{i 1} & \pi_{i 2} & \cdots & \pi_{i j} & \cdots\\
\cdots & \cdots & \cdots & \cdots & \cdots
\end{array}
\right)
\]
of non-negative integers $\pi_{i j}$ that are
non-increasing as
functions both of $i$ and $j$  \cite{macd}. The entries $\pi_{i j}$ are the parts of the
plane partition, and its norm (volume) is $|\bpi |=\sum_{i, j} \pi_{ij}$. Each plane partition is represented by a  three-dimensional Young
diagram consisting of unit cubes arranged into stacks so that the stack with coordinates $(i, j)$ is of height $\pi_{i j}$.
It is said that the plane partition is contained in a box with side
lengths $L, N, P$ if $i \leq L$, $j\leq N$ and $\pi_{ij} \leq P$ for all cubes of the Young diagram. A plane partition $\bpi$
that is decaying along each column and each row, $\pi_{i j}>\pi_{i+1, j}$ and $\pi_{i j} > \pi_{i, j+1}$, is called
\textit{strict plane partition} and denoted as $\bpi_{\rm spp}$. The element $(\bpi_{\rm spp})_{1 1}$ of strict plane partition
$\bpi_{\rm spp}$ satisfies the condition $(\bpi_{\rm spp})_{1 1} \geq 2N-2$ if all $i, j \leq N$. An arbitrary plane partition $\bpi$ in a box $N\times N\times P$ may be transferred into strict plane partition $\bpi_{\rm spp}$ in a box $N\times N\times (P+2N-2)$ by adding to $\bpi$ the matrix
\begin{equation}
{\bpi}_{\rm {min}} = \left(
\begin{array}{cccc}
2N-2 & 2N-3 & \cdots & N-1 \\
2N-3 & 2N-4 & \cdots & N-2 \\
\vdots & \vdots &  & \vdots \\
N-1 & N-2 & \cdots & 0
\end{array}
\right)
\label{min}
\end{equation}
which corresponds to a \textit{minimal} strict plane partition.

To connect the four-vertex model with plane partitions one may notice that
each admissible configuration of lattice paths may be associated
with an $N\times N$ array $\bpi_{\rm spp}$.
The $m^{\rm th}$ path (counting from the left) may be
thought of as the $m^{\rm th}$ column in this array with entries
$(\bpi_{\rm spp})_{km}$ equal to the number of the cells in the
subsequent columns $k$ of the lattice (starting from the right)
under the $m^{\rm th}$ path. The number of admissible configurations of lattice paths on a $2N \times (M+1)$ grid is equal to the number
of strict plane partitions in $N\times N \times M$ box.

The diagonals of an array $\bpi_{\rm spp}$ counting from the lower one (consisting of one element) are formed by the
partitions $\bmu^s$ with the elements equal to the number of cells lying under the horizontal parts of the paths in the $s^{\rm th}$ column ($s = 1, 2, \ldots, 2N-1$). The sum of the elements of the $s^{\rm th}$ diagonal is equal to the norm of the correspondent partition $|\bmu^s|=\sum_k \mu^s_k\,$, and the volume of the plane partition is $|\bpi_{\rm spp}| = |\bmu| = \sum_{s} | \bmu^s|$.
Let us introduce the notation $\tr_{\!s} \,\bpi_{\rm spp}$ for the sum of the entries of the matrix $\bpi_{\rm spp}$
along non-principal diagonals counted from the left down corner, we have $\tr_{\!s} \,\bpi_{\rm spp}
=|\bmu^s|$. The array
\begin{equation}
{\bpi}_{\rm spp} =\left(
\begin{array}{ccc}
6 & 5 & 3 \\
5 & 3 & 1 \\
4 & 1 & 0
\end{array}
\right)=
  \left(
\begin{array}{ccc}
\mu^3_1 & \mu^4_1 & \mu^5_1 \\
\mu^2_1 & \mu^3_2 & \mu^4_2 \\
\mu^1_1 & \mu^2_2 & \mu^3_3
\end{array}
\right)
\label{ar}
\end{equation}
corresponds to the nest of lattice paths in Fig.~6.
\begin{figure}[h]
\centering
\includegraphics{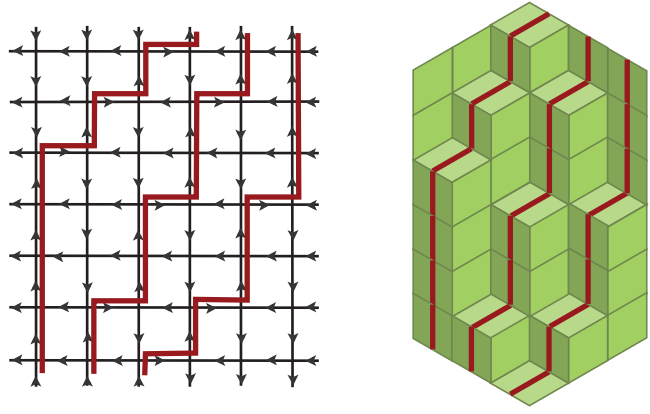}
\caption{A nest of admissible lattice paths with
the fixed boundary conditions and correspondent strict
plane partition.}
\end{figure}

The modified partition function (\ref{gfuhom}) may be expressed in the form
\begin{equation}\label{gfuhomp}
\widetilde{Z}_{\cal H} (e^{2{\bf h}_{\rm hom}}) = \sum_{\{\bpi_{\rm spp}\}} e^{2h|\bpi_{\rm spp}|} \,,
\end{equation}
where the sum
is taken over all strict plane partitions in $N\times N \times M$ box, and may
be considered as a partition function of strict three-dimensional Young diagram, where $2h$ plays the role of the chemical potential.
On the other hand, this partition function gives the generating function of strict plane partitions $G^{\,{\rm spp}}(N, N, M | a)$:
\begin{equation}
\label{gfssp}
\widetilde{Z}_{\cal H} ({\bf a}_{\rm hom})\,=\, G^{\,{\rm spp}} (N,N,M | a)\equiv\sum_{\{\bpi_{\rm spp}\}} a^{|\bpi_{\rm spp}|}\,,
\end{equation}
where $(2N-1)$-tuple ${\bf a}_{\rm hom}\equiv (a, a, \ldots, a)$ is defined, ${\bf a}_{\rm hom} = e^{2{\bf h}_{\rm hom}}$, i.e.,
$a=e^{2h}$. The number of strict plane partitions in $N\times N \times M$ box is obtained at $a\rightarrow 1$ and is given by
Eq.~(\ref{nspp}).

It is of interest to consider the weight $\bom$ in the $q$-parametrization:
\begin{equation}
\begin{array}{l}
\omega_j = q^{j/2}\,, \quad \omega_{N+j} = q^{(1-j)/2}\,,\qquad j=1, 2, \ldots, N\,,\\
[0.3cm]
{\bom} =
(\omega_1, \omega_2,
\ldots, \omega_{2N}) = {\bf q}^{\vee}\,,
\end{array}
\label{qp}
\end{equation}
where $2N$-tuple ${\bf q}^{\vee}$ is introduced:
\begin{equation}
\label{qp11}
{\bf{q}}^{\vee} \equiv (q^{1/2}, q, \ldots, q^{N/2}, 1, q^{-1/2},
\ldots, q^{-(N-1)/2})\,.
\end{equation}
It may be proved, \cite{bmumn, bt}, that the partition function (\ref{gfu1}) under this parametrization is the generating function of strict plane partition:
\begin{equation}
\widetilde{Z}_{\cal V} (\mathbf{q}^{\vee}) = q^{-\frac{N^2M}{2}}
\sum_{\{\bpi_{\rm spp}\}} q^{|\bpi_{\rm spp}|} =
q^{-\frac{N^2M}{2}}
\widetilde{Z}_{\cal H} ({\bf q}_{\rm hom})\,,
\label{pfgf}
\end{equation}
where the ``horizontal'' partition function (\ref{gfuhom})
depends on $(2N-1)$-tuple ${\bf q}_{\rm hom}\equiv (q, q, \ldots, q)$ introduced so that ${\bf q}_{\rm hom} = e^{2 {\bf h}_{\rm hom}}$.

The generating function of strict plane partitions with the fixed values of its diagonal parts is written in the form:
\begin{multline}\label{gffv}
G^{\,{\rm spp}} (N, N, M| q, \mathbf{a})=q^{\frac{N^2M}{2}}
\widetilde{Z}(\mathbf{q}^{\vee},
\mathbf{a})\\
=\sum_{\{\bpi_{\rm spp}\}} q^{|\bpi_{\rm spp}|}\prod_{s=1}^{2N-1} a_s^{|\bmu^s|}
\equiv \sum_{\{\bpi_{\rm spp}\}} q^{|\bpi_{\rm spp}|}\prod_{s=1}^{2N-1} a_s^{
\tr_{\!s} \,\bpi_{\rm spp}}\,,
\end{multline}
where $(2N-1)$-tuple  $\mathbf{a}=(a_1, a_2,\ldots , a_{2N-1})$ is introduced.

The volume of the plane partition (\ref{min}) is  equal to
\begin{equation}\label{volmin}
  |\bpi_{\rm min}|=\sum_{j,k=1}^N (\bpi_{\rm min})_{jk}= \sum_{j,k=1}^N\left( 2N-(j+k) \right)=N^2(N-1)\,,
\end{equation}
and therefore the
volumes of plane partitions $\bpi$ and of strict plane partitions $\bpi_{\rm spp}$ are related:
\begin{equation}\label{conn1}
|\bpi_{\rm spp}|\,=\, |\bpi| + N^2 (N-1)\,.
\end{equation}
The expression (\ref{conn1}) allow us to relate
the generating
functions of plane and strict plane partitions \cite{bmumn}:
\begin{multline}\label{conn}
G^{\,{\rm spp}}(N, N, M | q, \mathbf{a})\,=\,q^{N^2(N-1)} \\
\times \left(\prod_{j=1}^N a_j^{j} \prod_{j=1}^{N-1} a_{N+j}^{(N-j)} \right)^{N-1}\!\! G (N,N, M-2N+2 | q, \mathbf{a})\,.
\end{multline}
Taking into account Eq.~(\ref{gfua}), we obtain the explicit answer for the generating function of plane partitions in $N\times N\times P$ box with the fixed values of its diagonal parts:
\begin{multline}\label{gfuagf}
G (N,N,P | q, \mathbf{a})\\
=\,\prod_{1\leq m<k\leq N}\left(q^{k-1} \prod_{j=1}^{k-1} a_{N+j}-q^{m-1} \prod_{j=1}^{m-1} a_{N+j}\right)^{-1}
\prod_{1\leq m<k\leq N} \left(q^{k} \prod_{j=k}^{N}a_{j}-q^{m} \prod_{j=m}^{N}a_{j}\right)^{-1}\\
\times\,\det \left[\frac{1- \left(q^{k+m-1} \prod_{j=1}^{k-1} a_{N+j} \prod_{j=m}^N a_j \right)^{P+N-2}}
{1-q^{k+m-1} \prod_{j=1}^{k-1} a_{N+j}\prod_{j=m}^Na_j} \right]_{1\leq k,m \leq N}\,.
\end{multline}
In a box of  unbounded height $P$ and $N$ finite, the determinant in (\ref{gfuagf}) turns into the Cauchy determinant, and we obtain the answer:
\begin{equation}\label{infgf}
{\cal{G} }_N (q, \mathbf{a}) \equiv \lim_{P\to \infty}
G (N,N,P | q, \mathbf{a})
=
\prod_{k =1}^N \prod_{m=1}^N \frac{1}{1-q^{k+m-1} \prod_{j=1}^{k-1}a_{N+j} \prod_{j=m}^N a_j}\,.
\end{equation}

In the special case of inhomogeneous $(2N-1)$-tuple
\begin{equation}
\label{param1}
{\bf a} = (\underbrace{1, 1, \ldots, 1}_{s-1 {\,\,\rm times}}, {a},\underbrace{1, 1, \ldots, 1}_{2N-s-1 {\,\,\rm times}})\,,
\end{equation}
$1\le s\le N$, the relation (\ref{infgf}) transfers into
\begin{equation}\label{infgf1}
{\cal{G} }_N (q, \mathbf{a}) =
\prod_{k =1}^N \Bigg\{\prod_{m=1}^s \frac{1}{1- a q^{k+m-1}}
\,\prod_{m=s+1}^N \frac{1}{1-q^{k+m-1}}
\Bigg\}\,.
\end{equation}
Equation (\ref{infgf1}) is the generating function of plane partitions with fixed trace $\tr_{\!s}\,\bpi_{\rm}$,
$1\le s\le N$. Equation (\ref{infgf1}) at $s=N$ reduces to the norm-trace generating function obtained by Stanley in \cite{st}.

Let us introduce the $q$-deformed Barnes $G$-function:
\begin{equation}
G_q(n+1)\equiv \prod_{k=1}^{n}\Gamma_q (k)\,,
\label{gamma1}
\end{equation}
where $\Gamma_q(n)$ is $q$-Gamma function  \cite{kvant}:
\begin{equation}
\label{gamma}
\Gamma_q(n)\equiv [1]\,[2]\,\cdots\,[n-1]\,,\qquad
[k]\,\equiv\,\frac{1-q^k}{1-q}\,,\quad k\in {\BZ}^+\,.
\end{equation}
With the help of (\ref{gamma1}) and (\ref{gamma}) the generating function (\ref{infgf1}) may be brought into the form
\begin{align}
&{\cal{G} }_N (q, \mathbf{a}) =
(1-q)^{N(s-N)}\,{\cal{G} }^{\rm 1}_{N,s} (q)\, {\cal{G} }^{\rm 2}_N (q, a)\,, \label{gammma} \\
\label{gammma4}
&{\cal{G} }^{\rm 1}_{N,s} (q)
\equiv \frac{G_q(N+1) G_q(N+s+1)}{G_q(s+1) G_q(2N+1)}
\,,\qquad {\cal{G} }^{\rm 2}_N (q, a) \equiv \prod_{k=1}^N \frac{
(a, q)_{k}}{(a, q)_{k+s}}\,,
\end{align}
where $(a, q)_n$ is the shifted $q$-factorial:
\begin{equation}
\label{gammma1}
(a, q)_n \equiv (1-a)(1-aq)(1-aq^2)\cdots (1-a q^{n-1})\,,\qquad (a, q)_0=1\,.
\end{equation}
The function ${\cal{G} }^{\rm 1}_{N,s} (q)$ (\ref{gammma4}) is unity for $s=N$, and
Eq.~(\ref{gammma}) is simplified:
\begin{equation}
\label{gammma2}
{\cal{G} }_N (q, \mathbf{a})\, =\,\prod_{k=1}^N \frac{
(a, q)_{k}}{(a, q)_{k+N}}\,.
\end{equation}

Let ${\cal{G} }^{\rm spp}_N (q, \mathbf{a})$ be the generating function of strict plane partitions in box of infinite height $M\to\infty$ defined by analogy with (\ref{infgf}):
\begin{equation}
\label{gamm71}
{\cal{G} }^{\rm spp}_N (q, \mathbf{a}) \equiv \lim_{M\to \infty}
G^{\rm spp} (N,N,M | q, \mathbf{a})\,.
\end{equation}
In the case of the parametrization
(\ref{param1}) the definitions
(\ref{gammma}) and (\ref{gammma4}) are valid, and we obtain from (\ref{conn}):
\begin{equation}
{\cal{G} }^{\rm spp}_N (q, \mathbf{a}) = q^{N^2(N-1)} a^{s(N-1)}
(1-q)^{N(s-N)}\,{\cal{G} }^{\rm 1}_{N,s} (q)\, {\cal{G} }^{\rm 2}_N (q, a)\label{gamm72}\,.
\end{equation}

The choice of inhomogeneous $(2N-1)$-tuple
\[
{\bf a}\equiv (\underbrace{1, 1, \ldots, 1}_{N-l {\,\,\rm times}}, \underbrace{a, a, \ldots, a}_{l {\,\,\rm times}},\underbrace{1, 1, \ldots, 1}_{N-1 {\,\,\rm times}})
\]
leads to the generating function of plane partitions with the fixed value of the sum
$\sum_{s=N-l+1}^N \tr_{\!s}\,\bpi_{\rm}$, i.e.,
to an $l$-trace type formula of Gansner \cite{gan}:
\begin{equation}
\label{calg}
{\cal{G} }_N (q, \mathbf{a})=\prod_{k =1}^N \Bigg\{\prod_{m=1}^{N-l+1} \frac{1}{1- a^l q^{k+m-1}}
\,\prod_{m=N-l+2}^N \frac{1}{1- a^{N-m+1} q^{k+m-1}}
\Bigg\}\,.
\end{equation}
Equation (\ref{calg}) can be  also rewritten by means of (\ref{gammma1}):
\begin{align}
\nonumber
{\cal{G} }_N (q, \mathbf{a}) =
\prod_{m=1}^{N-l+1} \left(\frac{
(a^{N-m+1}, q)_{m+N}}{(a^{N-m+1}, q)_{m}}\,\frac{
(a^{l}, q)_{m}}{(a^{l}, q)_{m+N}} \right) &
\\
\times\,\prod_{k=1}^N \frac{
(a^{N-k+1}, q)_{k}}{(a^{N-k+1}, q)_{k+N}}\,.&
\label{gammma3}
\end{align}
Equation (\ref{gammma3}) is reduced to (\ref{gammma2}) at $l=1$.

\section{Conclusion}

Let us point out the new results obtained. The partition function of the anisotropic four-vertex model in the inhomogeneous external field is calculated. The derivation of the determinant representation of the partition function is based
on the QISM approach. The established connection of the considered model with the boxed plane partitions with the fixed sums of its
diagonals allows us to study their trace statistics. The obtained results are the generalization of the approach used in the paper \cite{bmnt} for the
calculation of the temporal evolution of the first moment of particles distribution of the quantum phase model. In the present paper we have concentrated on the calculation of the partition function of the
model in question. The problems of the trace statistic in the generic
case of an asymmetric box and the arctic curves phenomena are connected with calculation
of the correlation functions of the four-vertex model in presence of the external horizontal field.

\section{Acknowledgments}

We would like to thank A.~Pronko for valuable discussions. The work was supported by the Russian Science Foundation (Grant 18-11-00297).


\newpage
\begin{thebibliography}{99}

\bibitem{kor} V. E. Korepin, \textit{Calculation of norms of Bethe wave functions}, Comm. Math. Phys. \textbf{86}, 391 (1982).

\bibitem{iz} A. G. Izergin, \textit{Partition function of the six-vertex model in the finite volume}, Sov. Phys.
Dokl. \textbf{32},  878 (1987).

\bibitem{z} P. Zinn-Justin, \textit{Six-vertex model with domain wall boundary conditions and one-matrix
    model}, Phys. Rev. E \textbf{62}, 3411 (2000).

\bibitem{bpz} N. Bogoliubov, A. Pronko, M. Zvonarev, \textit{Boundary correlation functions of the six-vertex
    model}, J. Phys. A: Math. Gen. \textbf{35}, 5525 (2002).

\bibitem{bofv} N. M. Bogolyubov,  \textit{Five-vertex model with fixed boundary conditions}, St. Petersburg Math. J. \textbf{21}, 407
    (2010).

 \bibitem{cp1} F. Colomo, A. G. Pronko, \textit{The arctic curve of the domain-wall six-vertex model}, J. Stat. Phys. \textbf{138}, 662 (2010).

\bibitem{kp}  A. V. Kitaev, A. G. Pronko, \textit{Emptiness formation probability of the six-vertex model and the sixth Painlevé equation},
 Comm. Math. Phys. \textbf{345}, 305 (2016).

\bibitem{stan1}
           R.~Stanley, \textit{Enumerative combinatorics}, Vols. 1, 2, (Cambridge University Press, Cambridge, 1996, 1999).

\bibitem{kup}  G. Kuperberg, \textit{Another proof of the alternating-sign
matrix conjecture}, Int. Math. Res. Not. \textbf{1996}, 139
(1996).

\bibitem{bres}  D.~M. Bressoud, \textit{Proofs and Confirmations. The Story
of the Alternating Sign Matrix Conjecture}, (Cambridge University Press, Cambridge,
 1999).

\bibitem{bmumn} N.~M.~Bogoliubov, C.~Malyshev, \textit{Integrable models and combinatorics},
Russian Math. Surveys {\bf 70}, 789 (2015).

\bibitem{macd}  I. G. Macdonald, \textit{Symmetric functions and Hall
polynomials}, (Clarendon Press, 1995).

\bibitem{koh} K. Motegi, K. Sakai,
\textit{Vertex models, TASEP and Grothendieck polynomials},
J. Phys. A: Math. Theor. \textbf{46},  355201 (2013).

 \bibitem{resh} N. Reshetikhin, A. Sridhar,  \textit{Integrability of limit shapes of the six vertex model},
    Comm. Math. Phys. \textbf{56}, 535 (2017).

\bibitem{lyb} I. Lyberg, V. Korepin, G. Ribeiro, J. Viti, \textit{Phase separation in the six-vertex model with
a variety of boundary conditions}, J. Math. Phys. \textbf{59},  053301 (2018).

\bibitem{cp3} F. Colomo, A. Sportiello, \textit{Arctic curves of the six-vertex model on generic domains: the Tangent Method},
    J. Stat. Phys. \textbf{164}, 1488 (2016).

\bibitem{cp2} F. Colomo, A. G. Pronko, A. Sportiello, \textit{Arctic curve of the free-fermion six-vertex model in an L-shaped domain},
    J. Stat. Phys. \textbf{174}, 1 (2019).

\bibitem{gier} J. de Gier, R. Kenyon, S. S. Watson, \textit{Limit shapes for the asymmetric five vertex model}, arXiv:1812.11934.


\bibitem{bax} R. G. Baxter, \textit{Exactly Solved Models in Statistical
Mechanics} (San Diego, Academic press, 1982).

\bibitem{fad}  L. D. Faddeev, \textit{Quantum Inverse Scattering Method},
Sov. Sci. Rev. Math. \textbf{C1}, 107 (1980).

\bibitem{kbi}  V. E. Korepin, N. M. Bogoliubov, A. G. Izergin, \textit{Quantum Inverse Scattering Method and Correlation Functions}
(Cambridge University Press, Cambridge, 1993).

\bibitem{bt}
       N. M. Bogoliubov,
       \textit{Four-vertex model and random tilings},
Theor. Math. Phys. {\bf 155}, 523 (2008).

\bibitem{bj}
        N. M. Bogoliubov,
\textit{Four vertex model},
J. Math. Sci. (New York) {\bf 151}, 2816 (2008).


\bibitem{st} R. P.~Stanley, \textit{The conjugate trace and trace of a plane partition}, J. Comb. Theor. A \textbf{14}, 53 (1973).

\bibitem{gan} E. Gansner, \textit{The enumeration of plane partitions via the Burge correspondence}, Illinois J. Math. \textbf{25}, 533 (1981).

\bibitem{kam} S.~Kamioka, \textit{Plane partitions with bounded size of parts and biorthogonal polynomials}, arXiv:1508.01674.

\bibitem{tri} T.~Lai, \textit{A New Proof for a Triple Product Formula for Plane Partitions}, arXiv:1710.02241.

\bibitem{bmnt} N. M. Bogoliubov,  C. Malyshev, \textit{The phase model and the norm-trace generating function of plane partitions},
J. Stat. Mech.: Theory and Experiment, 083101 (2018).

\bibitem{kvant} A. Klimyk, K. Schmudgen, \textit{Quantum Groups and their Representations} (Springer-Verlag, Berlin Heidelberg, 1997)

\end{thebibliography}
\end{document}